%% file: paperITT_StructureCR_arXiv.tex
\documentclass[a4paper, 12pt]{article}
\usepackage{amsfonts, amsmath, amssymb, graphicx, color}
\usepackage[american]{babel}
\usepackage{latexsym}
\usepackage{epsfig}
\usepackage{theorem}

\newcommand{\exclusion}{[0,1]\setminus\{0,1/2,1 \}}
\newcommand{\pxone}{P_{\!\scriptscriptstyle{X_1}}}
\newcommand{\pxtwo}{P_{\!\scriptscriptstyle{X_2}}}

\newcommand{\defined}{{\stackrel{\scriptscriptstyle{\triangle}}{=}}}
\newcommand{\cX}{{\cal X}}
\newcommand{\cY}{{\cal Y}}

\newcommand{\cS}{{\cal S}}

\newcommand{\stir}[2]{{#1 \atopwithdelims\{\} #2}}
\newcommand{\coc}{\supset}
\newcommand{\QED}{\hspace{\fill}$\Box$}
\newcommand{\eqa}{\begin{eqnarray}}
\newcommand{\ena}{\end{eqnarray}}
\newcommand{\n}{\nonumber}

\newtheorem{definition}{Definition}
\newtheorem{theorem}[definition]{Theorem}
\newtheorem{proposition}[definition]{Proposition}
\newtheorem{lemma}[definition]{Lemma}
\newtheorem{corollary}[definition]{Corollary}
\newcommand{\btheorem}{\begin{theorem}\rm}
\newcommand{\etheorem}{\end{theorem}}
\newcommand{\bcorollary}{\begin{corollary}\rm}
\newcommand{\ecorollary}{\end{corollary}}
\newcommand{\blemma}{\begin{lemma}\rm}
\newcommand{\elemma}{\end{lemma}}
\newcommand{\bdefinition}{\begin{definition}\rm}
\newcommand{\edefinition}{\end{definition}}
\newcommand{\bproposition}{\begin{proposition}\rm}
\newcommand{\eproposition}{\end{proposition}}
\newcommand{\bproof}{\begin{proof}\rm}
\newcommand{\eproof}{\end{proof}}

\newcommand{\mR}{\mathcal{R}}
\newcommand{\mA}{\mathcal{A}}
\newcommand{\mB}{\mathcal{B}}
\newcommand{\mD}{\mathcal{D}}
\newcommand{\mT}{\mathcal{T}}
\newcommand{\mS}{\mathcal{S}}
\newcommand{\mK}{\mathcal{K}}
\newcommand{\mU}{\mathcal{U}}
\newcommand{\mV}{\mathcal{V}}

\newcommand{\mL}{\mathcal{L}}
\newcommand{\mQ}{\mathcal{Q}}
\newcommand{\mF}{\mathcal{F}}

\DeclareMathOperator{\coeff}{coeff}

\begin{document}

\title       {On the Structure of the Capacity Region of Asynchronous Memoryless Multiple-Access Channels}

\author      {Ninoslav Marina\thanks{N. Marina was with the School of Computer and Communication Sciences of the \'Ecole Polytechnique F\'ed\'erale de
Lausanne, CH-1015 Lausanne, Switzerland. He is now with the
Department of Electrical Engineering, University of Hawai`i at
M$\bar{\text{a}}$noa, Honolulu, HI 96822, USA. This work was partially supported by the Swiss National Science Foundation Grant Nr. 21-055699.98. The material
in this paper was presented in part at the IEEE International Symposium on
Information Theory, Chicago, USA, June/July 2004 \cite{ninoitt}.}~ and Bixio Rimoldi\thanks{B. Rimoldi is with
the School of Computer and Communication Sciences of the \'Ecole Polytechnique F\'ed\'erale de Lausanne, CH-1015
Lausanne, Switzerland.}}

\date{}
\maketitle

\begin{abstract}

The asynchronous capacity region of memoryless multiple-access channels is the union of certain polytopes. It is well-known that vertices of such polytopes may be approached  via a technique called successive decoding. It is also known that an extension of successive decoding applies to the dominant face of such polytopes.  The extension consists of forming groups of users in such a way that users within a group are decoded jointly whereas groups are decoded successively.  This paper goes one step further. It is shown that successive decoding extends to every face of the above mentioned polytopes. The group composition as well as the decoding order for all rates on a face of interest are obtained from a label assigned to that face. From  the label one can extract a number of structural properties, such as the dimension of the corresponding face and whether or not two faces intersect. Expressions for the the number of faces of any given dimension are also derived from the labels.

\vspace{2mm}


\end{abstract}

\section{Introduction}\label{intro}

The asynchronous capacity region of an $M$-user  memoryless multiple-access channel (MAC) is the union of certain $M$-dimensional polytopes. It is well known that if a desired rate tuple lies on the {\em vertex} of the so-called {\em dominant face} of such a polytope, one can decode {\em one user at a time successively}, using the codewords of already decoded users as side information \cite[Section 14.3.2]{cover}. For example,  for a 2-user code of blocklength $n$ and rates $R_1$ and $R_2$, respectively, decoding user 1 and 2 successively requires finding the first codeword  within a codebook of size $2^{nR_1}$ and subsequently finding the second codeword in a codebook of size $2^{nR_2}$. A decoder that makes a joint search does so in the bigger space of $2^{nR_1}2^{nR_2}$ pairs of alternatives. Hence the attractiveness of  successive decoding.

In \cite{gts} it is shown that successive decoding for dominant-face vertices extends to {\em group} successive  decoding for rate tuples that are in the boundary of the dominant face. More specifically, each point on the boundary of the dominant face is on a face of some dimension $k\in\{0,1,\dots,M-2\}$. For a rate tuple on such a face of dimension $k$,  successive decoding requires forming $M-k$ groups. For instance, for a vertex (a face of dimension $0$) we need $M$ groups,  which means that each ``group" contains a single user, implying, as it should, single user decoding of vertices. Alternatively, if the rate of interest  is on a face of dimension $1$, the number of groups is $M-1$, i.e., all  except  two
users can be decoded one at a time successively, and the group of two is decoded jointly.  In \cite{gts}  it is also shown that if the rate tuple of  interest is on the dominant face but not on its boundary, then one can split a user and a channel input and make sure that the new rate tuple, which has an additional component, lies on the boundary of the dominant face of the newly created channel. By iterating this procedure one obtains rate splitting multiple-access \cite{gruw,rim}.

In this paper we focus on some structural and operational properties of the $M$-dimensional polytopes that form the capacity region. We extend the labeling technique of \cite{ru99,gts} so as to have a label for every face. The label is unique if the polytope is non-degenerated. A degenerated polytope (to be properly defined later) is one for which certain faces collapse. To avoid complications due to the collapsing of faces we consider only non-degenerated cases. From the label, we can deduce structural properties such as which faces intersect and the dimensionality of a face. The label also specifies how to do  successive decoding of groups, which is an operational property. In particular, we will see that group decoding applies to every face (not only the faces of the dominant face).

The  paper is organized as follows. In Section \ref{labfac} we define the relevant polytopes and characterize and label their faces. The main result of Section  \ref{labfac}  is Proposition \ref{intersection}. It specifies which faces intersect and which do not. In Section \ref{sucdec} we make the link between the label and group successive decoding. In Section \ref{numdfac} we   give expressions for the number of faces of any given dimension. Section \ref{con} concludes the paper.

\section{Labeling faces}
\label{labfac}

Recall that an $M$-user discrete memoryless multiple-access channel
is defined in terms of $M$ discrete input-alphabets\footnote{All
results presented in this paper carry over to the Gaussian multiple-access channel.} ${\cal X}_i$, $i \in \{1, \cdots, M\}$, an output
alphabet ${\cal Y}$, and a stochastic matrix
$W:\cX_1\times\cX_2\times\cdots\times\cX_M \rightarrow \cY$ with
entries $W_{Y|X_1,X_2,\cdots,X_M}(y|x_1,x_2,\cdots,x_M)$ describing
the probability that the channel output is $y$ when the inputs are
$x_1,x_2,\cdots,x_M$. For any input distribution in product
form\footnote{Random variables and their sample values will be
  represented by capital and lowercase letters, respectively.}
$P_{X_1}, \cdots, P_{X_M}$, define ${\cal R}$ to be
\begin{equation} \label{equ:constraints}
{\cal R}= \{ R\in \mathbb{R}_+^M: R ({\cal S}) \leq I(X_{{\cal S}};
Y |X_{{\cal S}^c}), \quad \forall {\cal S} \subseteq [M]\},\n
\end{equation}
where $R(\mS)\defined\sum_{i\in\mS}R_i$, $X_{\cS} \defined (X_i)_{i\in\cS}$, $\cS^c  \defined
[M]\setminus\cS$, $[M]=\{1,2,\dots,M\}$, and $I(X_{{\cal S}}; Y |X_{{\cal S}^c})$ is the
mutual information between $X_{\cal S}$ and  $Y$ given  $X_{{\cal
S}^c}$. $\mathbb{R}_+$ denotes the nonnegative reals.  The capacity
region depends on whether or not there is synchronism. A
discrete-time
channel
is {\em synchronous} if the transmitters are able to index channel input
sequences in such a way that all inputs with time index $n$ enter
the channel at the same time.  If this is not the case, meaning that
there is an unknown shift between time indices, then the channel is
said to be {\em asynchronous.}

The capacity region for either the synchronous or asynchronous channel
may be described in terms of the region
\begin{equation} \label{equ:achievable}
{\cal C}_{DMC} = \bigcup_{\pxone\pxtwo\cdots P_{X_M}}
 {\cal R}[W; \pxone\pxtwo\cdots P_{X_M}],\n
\end{equation}
where the union is over all product input distributions.  The
capacity region of the asynchronous multiple-access channel with
arbitrarily large shifts between time indices is ${\cal C}_{DMC}$
\cite{Pol83,HuH85}, whereas if shifts are bounded or the
multiple-access channel is synchronous then its capacity region is
the convex hull of ${\cal C}_{DMC}$ \cite{ahlsw,liao,CMP81}.

\bdefinition \label{degeneratedCond} A region ${\cal R}$ is called non-degenerated if the following two conditions hold
\begin{enumerate}
\item \label{degeneratedCondOne} $I(X_{\mS};Y)>0$ for all non-empty sets $\mS\subseteq [M]$,
\item \label{degeneratedCondTwo} $I(X_{\mS};Y|X_{\mA})<I(X_{\mS};Y|X_{\mB})$ for all
$\emptyset\subset\mS\subset[M]$, $\mA\subset \mB \subset [M]$, and
$\mS\cap\mB=\emptyset$.
\end{enumerate}
\edefinition

The above definition is natural. Essentially it says that each input carries information and all inputs interfere with one another.
Notice that  for a non-degenerated channel it is also true that for all $\mA\subset[M]$, $\emptyset\subset\mS\subset\mT\subseteq [M]$, and $\mA\cap\mT=\emptyset$,
\begin{equation}
I(X_{\mS};Y|X_{\mA})<I(X_{\mT};Y|X_{\mA}) .
\label{degeneratedCondThree}
\end{equation}
To see this, we first observe that the independence of the input random variables implies that $I(X_{\mS};Y|X_{\mA}) \geq I(X_{\mS};Y)$ whenever $\mS$ and $\mA$ do not intersect. Thus, condition (\ref{degeneratedCondOne}) implies $I(X_{\mS};Y| X_{\mA})>0$ for every non-empty subset $\mS$ of $[M]$ and every subset $\mA$ of $[M]$ that does not intersect with $\mS$. Now we can use the chain rule of mutual information to obtain
$
I(X_{\mT};Y|X_{\mA}) = I(X_{\mS};Y|X_{\mA}) + I(X_{\mT \backslash \mS};Y|X_{\mA\cup\mS}) > I(X_{\mS};Y|X_{\mA})$, where the inequality holds since the second term on its left must be positive.

An example of a channel that does not fulfill condition (\ref{degeneratedCondOne}) above is the two-user binary adder channel when the sum is modulo $2$ and the inputs are assigned uniform probability. In this  case condition  (\ref{degeneratedCondOne}) is violated since $I(X_i;Y)=0$ for $i=1,2$.  Then $\mR$ is a triangle as opposed to a pentagon. An example for which condition  (\ref{degeneratedCondTwo}) is not fulfilled is
when we have two parallel channels.  In this case condition
  (\ref{degeneratedCondTwo})  is violated since $I(X_1;Y|X_2)=I(X_1;Y)=1$.  The same is true if we swap $X_1$ and $X_2$.
  In this case $\mR$ is a rectangle.

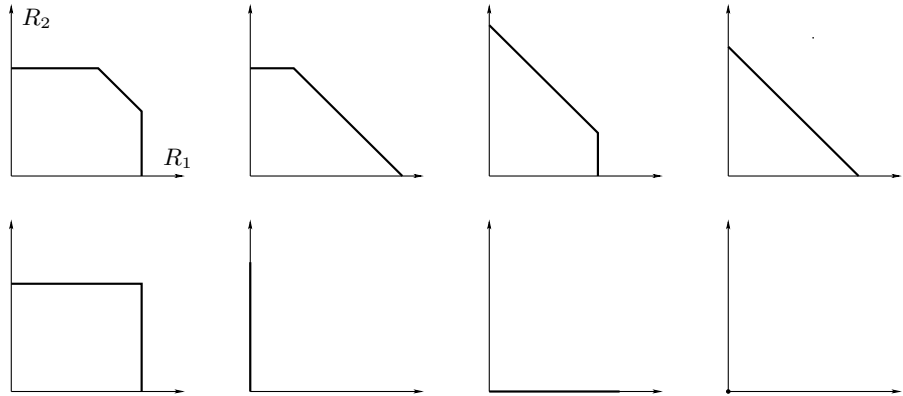
\begin{figure}[h]
\centering
\input{degenerated2.pstex_t}
\caption{Shapes of $\mR$ for a 2-user channel. The first sub-figure is that of a non-degenerated case. In all pictures, the abscissa represents $R_1$ and the ordinate $R_2$.} \label{dgn}
\end{figure}

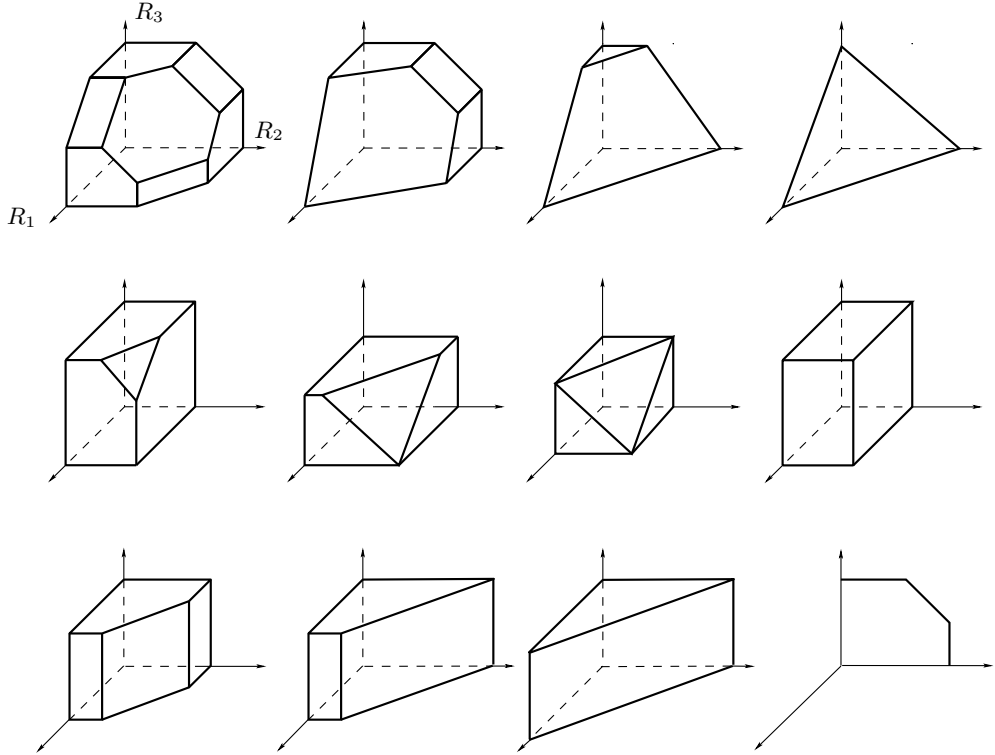
\begin{figure}[ht]
\centering
\input{cr3Ddegenerated2.pstex_t}
\caption{Some versions of $\mR$ for a 3-user channel. The first sub-figure is that of a non-degenerated case.}  \label{dgn3}
\end{figure}

Fig.~\ref{dgn} shows an example of a non-degenerated $\mR$ (first subfigure)  for $M=2$  and all possible degenerated variations.  Fig.~\ref{dgn3} shows examples of degenerated cases for $M=3$. All examples of  Fig.~\ref{dgn3} are for binary input channels and modulo $2$ sums (when applicable). The first row depicts regions for the channel $Y=X_1 + X_2 +  X_3$. If we denote by $p_i$ the probability that $X_i=1$, $i=1,2,3$, then the first region (non degenerated) is obtained with $p_i\in \exclusion$, $i=1,2,3$, the second region in the same row may be obtained with $p_1=0.5$, $p_2, p_3 \in  \exclusion$, the third with $p_1=p_2=0.5$, $p_3\in  \exclusion$, and the fourth with $p_1=p_2=p_3=0.5$. The first three subfigures of the second row correspond to  the channel $Y=
(Y_1,Y_2)=(X_1 +X_2, X_2 + X_3)$. The first region
 may be obtained with $p_i\in  \exclusion$, $i=1,2,3$, the second with $p_1=p_2=0.5$, $p_3\in  \exclusion$, and the third with $p_1=p_2=p_3=0.5$. The last region in the second row may be obtained from the MAC $Y=(Y_1,Y_2,Y_3)= (X_1,X_2,X_3)$ with $p_i\in \exclusion$, $i=1,2,3$. The first three subfigures in the third row correspond to the channel $Y=(Y_1,Y_2)=(X_1 +X_2, X_3)$, with the input distributions of the first one being
$p_1, p_2\in  \exclusion, p_3\in (0,1)$, of the second being $p_1\in \exclusion,p_2=0.5, p_3\in (0,1)$, and of the third being $p_1=p_2=0.5$, $p_3\in (0,1)$. The last figure in the third row may be obtained with $p_1\in \{0,1\}$ and $p_i\in (0,1)$, $i=2,3$.

An object of the form $\{R\in\mathbb{R}_+^M: R(\mS)=c\}$, for some
constant $c$, is an hyperplane of $\mathbb{R}_+^M$ of dimension
$M-1$. The set $\{R\in\mathbb{R}_+^M: R(\mS)\leq c\}$ is one of the
two half-spaces bounded by such an hyperplane. $\mR$ is a finite
intersection of such half-spaces.  A linear inequality $ Ra\leq
a_0$, where $R$ is  a row vector, $a$ a column vector and $a_0$ is a scalar, is valid
for $\mR$ if it is satisfied for all points $R\in \mR$. A {\it face}
of $\mR$ is defined as any set of the form 
\eqa \mF&=& \mR \cap\{R\in
\mathbb{R}_+^M: Ra=a_0\}, \n
\ena 
where $ Ra \leq a_0$ is a valid
inequality for $\mR$. The dimension of a face is the dimension of
its affine hull, namely dim$(\mF):=$dim(aff$(\mF)$). In words, a
face of $\mR$ is the intersection of  $\mR$ with an $(M-1)$
dimensional hyperplane that keeps  $\mR$ on one side. Since the
inequality $R{\mathbf 0}\leq 0$ (${\mathbf 0}$ being all zero
vector) is valid for $\mR$, we observe
that $\mR$ itself is a face. All the other faces $\mF$, called {\it proper} faces,
 satisfy $\mF\subset \mR$. Note that the number of faces of any dimension is maximal
 in the non-degenerated. In the following text we consider only channels with non-degenerated regions.

Faces of dimension $0,1$, $M-2$, and $M-1$ are called $vertices$,
$edges$, $ridges$, and $facets$, respectively. In the non-degenerated case,
for a single user channel, $\mR$
has two vertices and one edge and for a $2$-user channel it
has five vertices and five edges. In
Fig.~\ref{poly3D} we see that there are $16$ vertices, $24$ edges and
10 facets for a non-degenerated region $\mR$ of a $3$-user channel. 
 
For every $i\in [M]$, there is a {\em back facet} of the form
 \eqa \mB_i=\mR\cap \{R\in \mathbb{R}_+^M: R_i=0\},
\label{bf}\n\ena and for every
 $\mS \subseteq [M]$, $\mS\ne\emptyset$, there
is a {\em front facet} \eqa \mF_{\mS}=\mR\cap\{R\in \mathbb{R}_+^M:
R(\mS)=I(X_{\mS};Y|X_{\mS^c})\}\label{ff}.\n \ena
There are $M$ back facets and $2^M-1$ front facets, one for each
non-empty subset of $[M]$. It is convenient to extend the notation $\mB_i$ and $\mF_{\mS}$ as follows 
\begin{align*} \mB_{\mA}&=\bigcap_{i\in\mA}\mB_i, \, \text{ with } \mB_\emptyset=\mR \text{ by convention }, \\
\mF_{\mS_1,\mS_2,\ldots,\mS_m}&=\bigcap_{j=1}^m\mF_{\mS_j}, \, \text{ with } \mF_\emptyset=\mR \text{ by convention },\\
\mF_{\mS_1,\mS_2,\ldots,\mS_m|\mA}&= \mF_{\mS_1,\mS_2,\ldots,\mS_m} \cap\mB_{\mA}.
\end{align*} Note that $\mF_{\mS|\emptyset}=\mF_{\mS}$, $\mF_{\emptyset|\mA}=\mB_{\mA}$, and
$\mF_{\emptyset|\emptyset}=\mR$. Fig.~\ref{poly3D} shows a non-degenerated $\mR$ for a $3$-user channel and some of the labels.
\begin{figure}[h]
\centering
\input{cr3Dbest.pstex_t}
\caption{Region $\mR$ with labels for a $3$-user MAC.} \label{poly3D}\end{figure}

Next we show that two front facets intersect if and only if the index
set of one is a subset of the index set of the other
(Lemma \ref{lemmaff}) and that a front and a back facet intersect if
and only if the index of the back facet is not
an element of the set that defines the front facet (Lemma
\ref{lemmabf}).

\blemma \label{lemmaff}  $\mF_{\mS_1}\cap
\mF_{\mS_2}$ is not empty iff $\mS_1\subseteq \mS_2$ or $\mS_2\subseteq
\mS_1$. \elemma
{\it Proof:}
The ``if'' direction is clearly true if $\mS_1=\mS_2$. Assume without loss of generality
that $\mS_1\subset \mS_2$. We want to show the existence of an
$R\in\mR$ such that 
\eqa 
R(\mS_1)&=&I(X_{\mS_1};Y|X_{\mS_1^c})\label{one1},\text{  and}\n\\
R(\mS_2)&=&I(X_{\mS_2};Y|X_{\mS_2^c})\label{two2}. \n\ena

Without loss of generality, we re-index users so that $\mS_1=[k]$
and $\mS_2=[\ell]$, where $\ell>k$. Consider $R=(R_1,\dots,R_M)$
defined as follows
\eqa R_{i}=\begin{cases}I(X_{i};Y|X_{i+1},\dots,X_M), &
i=1,\dots,M-1,\\ I(X_{M};Y)&i=M.
\end{cases} \n\ena

Observe that $R$ is a vertex of the dominant face. Hence $R\in\mR$. Furthermore, from the chain rule for
mutual information \eqa
R([i])=\sum_{j=1}^iR_j=\sum_{j= 1}^i I(X_{j};Y|X_{j+1},\dots,X_M)=I(X_{[i]};Y|X_{[i]^c}).\n\ena

Thus, for $i=k$ we get $R([k])=I(X_{\mS_1};Y|X_{\mS_1^c})$ and
for $i=\ell$, $R([\ell])=I(X_{\mS_2};Y|X_{\mS_2^c})$. Hence $R\in
\mF_{\mS_1}\cap \mF_{\mS_2}$.

To prove the ``only if'' direction, let $R\in
\mF_{\mS_1}\cap\mF_{\mS_2}$. Then
\eqa I(X_{\mS_1\cup\mS_2};Y|X_{(\mS_1\cup\mS_2)^c})
&\stackrel{(a)}\geq&R(\mS_1\cup\mS_2)=R(\mS_1)+R(\mS_2)-R(\mS_1\cap\mS_2)\n\\
&\stackrel{(b)}=&I(X_{\mS_1};Y|X_{\mS_1^c})+I(X_{\mS_2};Y|X_{\mS_2^c})-R(\mS_1\cap\mS_2)\n\\
&\stackrel{(c)}\ge& I(X_{\mS_1};Y|X_{\mS_1^c})+I(X_{\mS_2};Y|X_{\mS_2^c}) - I(X_{\mS_1\cap\mS_2};Y|X_{(\mS_1\cap\mS_2)^c}) \n\\
&\stackrel{(d)}=&I(X_{\mS_1\setminus\mS_2};Y|X_{\mS_1^c})+I(X_{\mS_2};Y|X_{\mS_2^c})\n\\
&\stackrel{(e)}\ge&I(X_{\mS_1\setminus\mS_2};Y|X_{(\mS_1\cup\mS_2)^c})+I(X_{\mS_2};Y|X_{\mS_2^c})\n\\
&=& I(X_{\mS_1\cup\mS_2};Y|X_{(\mS_1\cup\mS_2)^c})
\label{contr}\n\ena where $(a)$ and $(c)$ follow from the fact that
$R\in\mR$,  $(b)$ from the definition of $\mF_{\mS_i}, i=1,2$, $(d)$
from the chain rule for mutual information, and $(e)$ holds since
the inputs are independent and  conditioning on independent inputs
can not decrease mutual information. By comparing the first and the
last term of the above chain, we see that (a), (c), and (e) must be equalities.  Equality in $(e)$ means
\eqa I(X_{\mS_1\setminus\mS_2};Y|X_{\mS_1^c})=I(X_{\mS_1\setminus\mS_2};Y|X_{(\mS_1\cup\mS_2)^c}).\n \ena
Since $\mR$ is non-degenerated (by assumption), the above equality implies that either $\mS_1 \setminus \mS_2 = 0$, i.e., $\mS_1 \subseteq \mS_2$ or $\mS_1 = \mS_1 \cup \mS_2$, i.e., $\mS_2 \subseteq \mS_1$.
This completes the proof. 

\QED


The following Lemma is from \cite{gts}.

\blemma \label{lemmabfgen} Assume $R\in\mF_{\mS}$. Then for every $\mL
\subseteq \mS$ \eqa I(X_{\mL};Y|X_{\mS^c})\le R(\mL) \le
I(X_{\mL};Y|X_{\mL^c}).\ena
 \elemma
{\it Proof:} The second inequality is true
for every $R\in\mR$. To prove the first inequality observe that \eqa
R(\mL)&=& R(\mS)-R(\mS\setminus\mL)\n\\
&\stackrel{(a)}=& I(X_{\mS};Y|X_{\mS^c})-R(\mS\setminus\mL)\n\\
&\stackrel{(b)}\ge&
I(X_{\mS};Y|X_{\mS^c})-I(X_{\mS\setminus\mL};Y|X_{(\mS\setminus\mL)^c})\n\\
&\stackrel{(c)}=&I(X_{\mL};Y|X_{\mS^c}),\ena where $(a)$ is true
since $R\in\mF_{\mS}$, $(b)$ since $R\in\mR$ and $(c)$ follows from
the chain rule for mutual information.

\QED

\blemma \label{lemmabf} ${\cal F}_{\mS}\cap\mB_\mA\ne\emptyset$
iff $\mA\cap \mS=\emptyset$. \elemma

{\it Proof:} If $\mA=\emptyset$ then the Lemma is clearly true. Assume $\mA\not=\emptyset$.  To prove one direction, let  and  $R\in \mF_{\mS}\cap \mB_\mA$. Then  $0=R(\mA)=R(\mS\cap\mA)\ge
I(X_{\mS\cap\mA};Y|X_{\mS^c})$, where the inequality follows  from Lemma
\ref{lemmabfgen}. This implies that $I(X_{\mS\cap\mA};Y|X_{\mS^c})=0$. Since $R$ is non-degenerated, it follows  that
$\mA\cap \mS=\emptyset$. To prove the other direction, assume $\mA\cap \mS=\emptyset$ and pick a rate $\tilde R$ such that  $\tilde R\in\mF_\mS$. Let $R$ be obtained from $\tilde R$ by setting to $0$ all coordinates with index in $\mA$. Clearly $R\in\mB_\mA$ but also $R\in\mF_\mS$ since $R(\mS)=\tilde R(\mS)$.

\QED

\bproposition \label{intersection} The intersection
$\mF_{\mS_1,\mS_2,\ldots,\mS_m|\mA}$ is not empty, if and only if
the following two conditions are satisfied 

(i)  The set sequence $\mS_1,\mS_2,\ldots,\mS_m$ is telescopic, i.e., there is a permutation $\pi$ on the index set $[m]$ such that $\mS_{\pi(1)}\coc\mS_{\pi(2)}\coc\dots\coc\mS_{\pi(m)}$, and

(ii)  $\mA\cap\mS_{\pi(1)}=\emptyset$. \eproposition

{\it Proof:} Assume that, after re-indexing if necessary,  $\mS_1\coc\mS_2\coc\dots\coc\mS_m$
and $\mA\cap\mS_1=\emptyset$. The construction in the ``if" part of the proof of Lemma \ref{lemmaff} leads to an $\tilde R$ in $\mF_{\mS_1,\mS_2,\ldots,\mS_m}$. Let $R$ be obtained
from $\tilde R$ by setting to $0$ all coordinates with index in $\mA$. This does not affect coordinates with index in $\mS_i$.  Hence $R\in\mF_{\mS_i}$,  $i=1,\dots,m$ and $R\in\mB_\mA$.

To prove the converse, we observe that if $\mS_i$ is not contained in  $\mS_j$ or vice versa,  then by Lemma \ref{lemmaff},
$\mF_{\mS_i}\cap\mF_{\mS_j}=\emptyset$. Similarly, if
$\mS_1\cap\mA\not=\emptyset$, then according to Lemma \ref{lemmabf},
$\mF_{\mS_1}\cap \mB_\mA = \emptyset$. This concludes the proof.  

\QED

Note that proposition \ref{intersection} allows us to define a unique label for each face in $\mR$.

\begin{figure}[ht]
\begin{center}
\includegraphics[height=10cm, width=15cm]{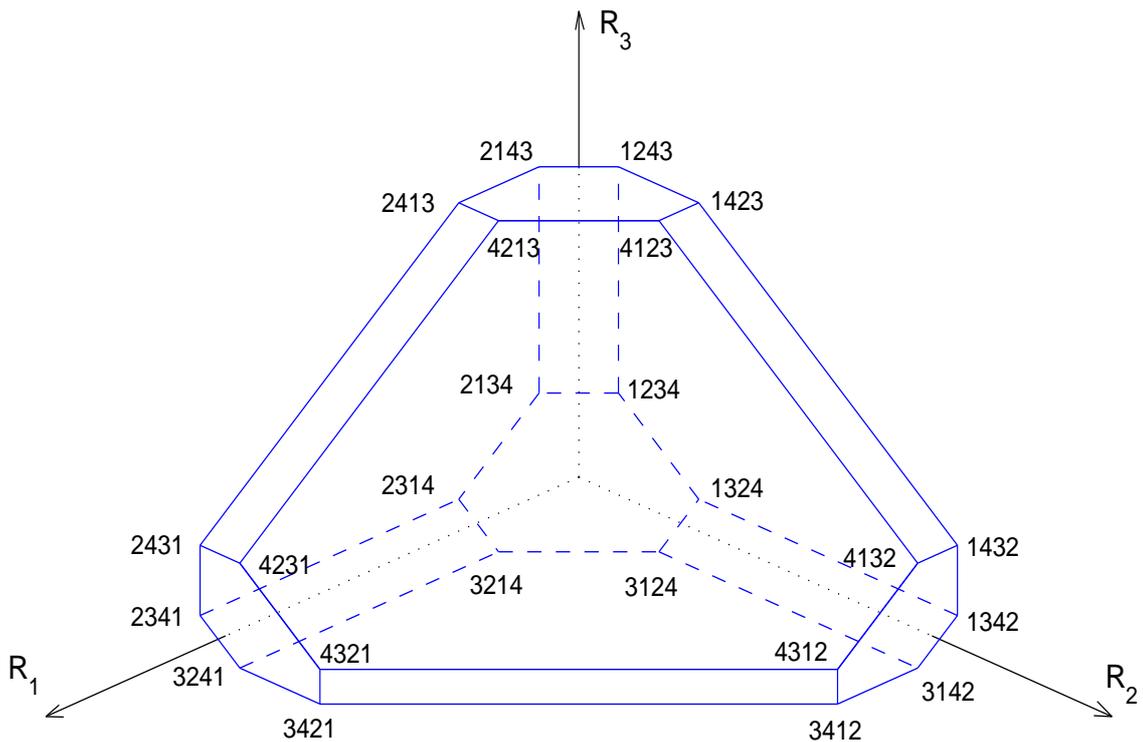}
\caption{Dominant facet of a $4$-user MAC. The $4$th dimension, not shown here, has coordinate
$R_4=I(X_{\{1,2,3,4\}};Y)-R_1-R_2-R_3$. Labels describe the decoding order used to approach the corresponding vertex via successive decoding.}
\end{center}
\label{4D}
\end{figure}

There is one facet of $\mR$ that stands out from the
others. It is the {\em dominant facet} (commonly called dominant
face) $\mF_{[M]}$. It is special since points in the dominant facet have maximal sum-rate. Observe that, from Lemma \ref{lemmabf}, the dominant facet
is the only facet that does not intersect with any back facet. The
structure of the dominant facet was presented in \cite{ru99}. In the
one-user case, the dominant facet is a vertex, in the two-user case
it is an edge that has two vertices, in the three-user case a
hexagon (Fig.~\ref{poly3D}). In
Fig.~\ref{4D}, we see that there are $24$ vertices, $36$ edges and $14$
two-dimensional faces in the dominant facet of a $4$-user channel. In general, the dominant facet is a
geometrical object called {\it permutahedron} \cite{zieg}. 
The notation for a vertex in Fig.~\ref{4D}, has been simplified.
Instead of writing the telescopic sequence
$\mF_{\{1,2,3,4\},\{1,2,3\},\{1,3\},\{3\}}$, we have written the
sequence of ``decrements,'' i.e., $4,2,1,3$ (commas
are not shown in Fig.~\ref{4D}). Besides being more compact,
the sequence of decrements gives the order in which users are decoded.
It is also a convenient notation to count vertices. Since each
permutation on the set $[M]$ is a vertex in the dominant facet, it
is clear that there are $M!$ such vertices.

\section{Structure, dimensionality, and group successive decoding}
\label{sucdec}
\newcommand{\tR}{\tilde{R}}
\newcommand{\dt}{\mD}
In this section we show that the faces  of $\mR$ consist of the Cartesian product of fundamental regions and of dominant facets of channels that are ``spin-offs" from the original channel $W$. To distinguish those channels,
we use subscripts that indicate the channel inputs and outputs. The
original channel $W$ will be denoted by $W_{Y|X_{[M]}}$. Recall that the
region $\mR$ is completely specified by the channel
$W_{Y|X_{[M]}}$ and by the input distribution
$P_{X_{[M]}}$.

For any two sets $\mU,\mV \subset [M]$ such that $\mU\cap\mV=\emptyset$,
there is a channel with inputs $X_{\mU}$ and outputs $(Y, X_{\mV})$. Specifically,
\eqa W_{YX_{\mV}|X_{\mU}}(y,x_{\mV}|x_{\mU})&=&
P_{X_{\mV}}(x_{\mV})W_{Y|X_{\mU},X_{\mV}}(y|x_{\mU},x_{\mV}))\n\\
&=&P_{X_{\mV}}(x_{\mV})\sum_{x_{[M]\setminus\mU\setminus\mV}}W_{YX_{[M]\setminus\mU\setminus\mV}|X_{\mU},X_{\mV}}(y,x_{[M]\setminus\mU\setminus\mV}|x_{\mU},x_{\mV})\n\\
&=&P_{X_{\mV}}(x_{\mV})\sum_{x_{[M]\setminus\mU\setminus\mV}}P_{X_{[M]\setminus\mU\setminus\mV}}(x_{[M]\setminus\mU\setminus\mV})W_{Y|X_{[M]}}(y|x_{[M]}),\n\ena
where by convention $P_{X_{\emptyset}}(x_{\emptyset})=1$.

A rate tuple for $W_{YX_{\mV}|X_{\mU}}$ is an expression of the form  $R_{\mU}\defined(R_i)_{i\in\mU}$. The corresponding fundamental region  $\mR_{YX_{\mV}|X_{\mU}}$ is defined by \eqa
\mR_{YX_{\mV}|X_{\mU}}\stackrel{\Delta} =
\mR[W_{YX_{\mV}|X_{\mU}};P_{X_{\mU}}] = \{R \in\mathbb{R}_+^{|\mU|}:
\quad R(\mL)\le I(X_{\mL};Y|X_{\mV\cup(\mU\setminus\mL)}), \quad
\forall\mL\subseteq \mU\}. \label{defr3}\ena The dimensionality
of $\mR_{YX_{\mV}|X_{\mU}}$ is $|\mU|$. Its dominant facet is the
$(|\mU|-1)$-dimensional subregion obtained by adding
the equality $R(\mU) = I(X_{\mU};Y|X_{\mV})$ i.e., \eqa \mD_{YX_{\mV}|X_{\mU}}&\stackrel{\Delta} =&
\mD[W_{YX_{\mV}|X_{\mU}};P_{X_{\mU}}] \n\\ &=&
\{R \in\mathbb{R}_+^{|\mU|}: R(\mL)\le
I(X_{\mL};Y|X_{\mV\cup(\mU\setminus\mL)}), \forall\mL\subset \mU,
R(\mU) = I(X_{\mU};Y|X_{\mV})\}. \label{defd3}\n\ena

The following special cases will be used frequently 
\eqa \mR_{YX_{\mS^c}|X_{\mS}}&\stackrel{\Delta} =&
\mR[W_{YX_{\mS^c}|X_{\mS}};P_{X_{\mS}}] =
\{R \in\mathbb{R}_+^{|\mS|}: \quad R(\mL)\le I(X_{\mL};Y|X_{\mL^c}),
\quad \forall\mL\subseteq \mS\},\n\\
\mD_{YX_{\mS^c}|X_{\mS}}&\stackrel{\Delta} =&
\mD[W_{YX_{\mS^c}|X_{\mS}};P_{X_{\mS}}] \n\\
&=& \{R \in\mathbb{R}_+^{|\mS|}: R(\mL)\le I(X_{\mL};Y|X_{\mL^c}),
\forall\mL\subset \mS, R(\mS) = I(X_{\mS};Y|X_{\mS^c})\}, \n\\
\mR_{Y|X_{\mS}}&\stackrel{\Delta} =& \mR[W_{Y|X_{\mS}};P_{X_{\mS}}]
= \{R \in\mathbb{R}_+^{|\mS|}: \quad R(\mL)\le
I(X_{\mL};Y|X_{\mS\setminus\mL}), \quad \forall\mL\subseteq
\mS\},\label{rdef2}\n\\
\mD_{Y|X_{\mS}}&\stackrel{\Delta} =&
\mD[W_{Y|X_{\mS}};P_{X_{\mS}}]\n\\
&=& \{R \in\mathbb{R}_+^{|\mS|}:  R(\mL)\le
I(X_{\mL};Y|X_{\mS\setminus\mL}), \forall\mL\subset \mS, R(\mS) =
I(X_{\mS};Y)\}. \label{defd2}\n\ena

The next lemma says that $\mF_{\mS}$ is the Cartesian product of a fundamental region $\mR$ and a dominant facet $\mD$. One expects this to be the case by looking at the facets $\mF_{\{1\}}$, $\mF_{\{2\}}$, and
$\mF_{\{3\}}$ of Fig. \ref{poly3D}. For instance $\mF_{\{1\}}$ is
the Cartesian product of a singleton and a pentagon. The singleton, the value of $R_1$, is the dominant facet $\mD$ of a single-user channel.
The pentagon, the region that contains $R_{\{2,3\}}$,  is the region $\mR$ of a two-user multiple-access channel. A perhaps less evident example is $\mF_{\{1,2\}}$. This is the Cartesian product of the dominant facet $\mD$ of a two user multiple-access channel and the fundamental region $\mR$ of a single user channel.

\blemma \label{theo4} $R\in\mF_{\mS}$ iff
$R_{\mS^c}\in\mR_{Y|X_{\mS^c}}$ and
$R_{\mS}\in\mD_{YX_{\mS^c}|X_{\mS}}$. \elemma

$Proof:$ Let $R\in \mF_{\mS}$. From the definition of $\mF_{\mS}$,
$\forall \mL\subset \mS\subseteq [M]$, $R(\mL)\le
I(X_{\mL};Y|X_{\mL^c})$ and $R(\mS)=I(X_{\mS};Y|X_{\mS^c})$.
Therefore $R_{\mS}\in \mD_{YX_{\mS^c}|X_{\mS}}$. Moreover, $\forall
\mT\subset\mS^c$ we may write $[M]=\mS\cup\mT\cup\mQ$ as the union
of disjoint sets. Then \eqa R(\mT)+R(\mS)&\le&
I(X_{\mT\cup\mS};Y|X_{\mQ})\n\\
&=&I(X_{\mT};Y|X_{\mQ})+I(X_{\mS};Y|X_{\mQ\cup\mT})\n\\
&=&I(X_{\mT};Y|X_{\mQ})+I(X_{\mS};Y|X_{\mS^c})\n\\
&=&I(X_{\mT};Y|X_{\mS^c\setminus\mT})+R(\mS).
 \n\ena
Hence $R(\mT)\le I(X_{\mT};Y|X_{\mS^c\setminus\mT})$ and from
(\ref{defr3}) it follows that $R_{\mS^c}\in\mR_{Y|X_{\mS^c}}$.

To prove the converse, let $R_{\mS}\in\mD_{YX_{\mS^c}|X_{\mS}}$ and
$R_{\mS^c}\in\mR_{Y|X_{\mS^c}}$. We have to prove that
$R(\mS)=I(X_{\mS};Y|X_{\mS^c})$ and that for all $\mL\subseteq [M]$,
$R(\mL)\le I(X_{\mL};Y|X_{\mL^c})$. The former is true since
$R_{\mS}\in\mD_{YX_{\mS^c}|X_{\mS}}$. To prove the latter, let
$\mT=\mL\cap\mS$ and $\mQ=\mL\cap\mS^c$. Since
$R_{\mS}\in\mD_{YX_{\mS^c}|X_{\mS}}$, $R(\mT)\le
I(X_{\mT};Y|X_{\mS^c\cup(\mS\setminus\mT)})=I(X_{\mT};Y|X_{\mT^c})$
for all $\mT\subseteq\mS$. Furthermore, since
$R_{\mS^c}\in\mR_{Y|X_{\mS^c}}$, $R(\mQ)\le
I(X_{\mQ};Y|X_{\mS^c\setminus\mQ})$ for all $\mQ\subseteq\mS^c$.
Hence \eqa
R(\mL)&=&R(\mT\cup\mQ)=R(\mT)+R(\mQ)\n\\
&\le& I(X_{\mT};Y|X_{\mT^c})+I(X_{\mQ};Y|X_{\mS^c\setminus\mQ})\n\\
&\le&
I(X_{\mT};Y|X_{\mT^c})+I(X_{\mQ};Y|X_{(\mS^c\setminus\mQ)\cup(\mS\setminus\mT)})\n\\
&=&I(X_{\mT\cup\mQ};Y|X_{(\mT\cup\mQ)^c})\n\\
&=&I(X_{\mL};Y|X_{\mL^c})\n\ena for all $\mL\subseteq [M]$ and this
completes the proof. 

\QED

From Lemma  \ref{theo4} we obtain the dimension of $\mF_{\mS}$ 
\eqa
\dim(\mF_{\mS})&=&\dim(\mR_{Y|X_{\mS^c}})+\dim(\mD_{YX_{\mS^c}|X_{\mS}})\n\\
&=&|S^c|+|S|-1=M-1,\n\ena
which is to be expected for a facet.

Finally, Lemma \ref{theo4} tells us that a rate point in $\mF_{\mS}$ may be approached via group successive decoding where groups are decoded in
the ord$(\mS^c,\mS)$. (For a rigorous proof of this fact we need to use codes that have independent and identically distributed components. This may be done using random coding arguments as in \cite{gts}.)

Then next result is a generalization of Lemma \ref{theo4}. It says that when $\mF_{\mS_1,\mS_2,...,\mS_m}$ is not empty it is the Cartesian product of a fundamental region and $m$ dominant facets.

\btheorem \label{theo6} Let $\mS_1\supset  \mS_2 \ldots \supset \mS_m$ form a telescopic sequence.
$R\in\mF_{\mS_1,\mS_2,...,\mS_m}$ iff
$R_{\mS_1^c}\in\mR_{Y|X_{\mS_1^c}}$ and $R_{\mS_i\setminus\mS_{i+1}}\in
\mD_{YX_{\mS_i^c}|X_{\mS_i\setminus\mS_{i+1}}}$ for $i=1,\dots,m$,
where by way of convention we  have defined $\mS_{m+1}=\emptyset$.
\etheorem

$Proof:$ Let $R \in \mF_{\mS_1, \ldots, \mS_m}$ and recall that
$\mF_{\mS_1,\mS_2,\dots,\mS_m}=\bigcap_{i=1}^m\mF_{\mS_i}$. From Lemma
\ref{theo4} we have \eqa R_{\mS_i}\in\mD_{YX_{\mS_i^c}|X_{\mS_i}}
\quad \text{and} \quad R_{\mS_i^c} \in \mR_{Y|X_{\mS_i^c}}, \quad
i=1,\dots,m. \label{AAA2}\ena This proves
$R_{\mS_1^c}\in\mR_{Y|X_{\mS_1^c}}$. In order to complete the proof
of the direct part, it is sufficient to show that (\ref{AAA2})
implies \eqa R_{\mS_i\setminus\mS_{i+1}}\in
\mD_{YX_{\mS_i^c}|X_{\mS_i\setminus\mS_{i+1}}}, \quad \forall
i=1,\dots, m-1. \n\ena For this, it is enough to show that,
$R(\mK)\le
I(X_{\mK};Y|X_{\mS_i^c\cup(\mS_i\setminus\mS_{i+1}\setminus\mK)})=I(X_{\mK};Y|X_{(\mS_{i+1}\cup\mK)^c})$
$\forall\mK\subseteq\mS_i\setminus\mS_{i+1}$, $i=1,\dots,m$, with
equality if $\mK=\mS_i\setminus\mS_{i+1}$. From (\ref{AAA2}), for any
$\mK\subseteq\mS_i$, $R(\mK)\le I(X_{\mK};Y|X_{\mK^c})$ with
equality if $\mK=\mS_i$ and for any $\mL\subseteq\mS_{i+1}$,
$R(\mL)\le I(X_{\mL};Y|X_{\mL^c})$ with equality if $\mL=\mS_{i+1}$.
Then for $\mK\subseteq\mS_i\setminus\mS_{i+1}$
we have \eqa R(\mK)&=&R(\mS_i)-R(\mS_{i+1})-R(\mS_i \setminus\mS_{i+1}\setminus\mK)\n\\
&=&I(X_{\mS_i};Y|X_{\mS_i^c})-I(X_{\mS_{i+1}};Y|X_{\mS_{i+1}^c})-R(\mS_i\setminus\mS_{i+1}\setminus\mK)\n\\
&\stackrel{(a)}{\le}&I(X_{\mS_i};Y|X_{\mS_i^c})-I(X_{\mS_{i+1}};Y|X_{\mS_{i+1}^c})-I(X_{\mS_i\setminus\mS_{i+1}\setminus\mK};Y|X_{\mS_i^c})\n\\
&=&I(X_{\mK};Y|X_{(\mS_{i+1}\cup\mK)^c}),\n\label{equa} \ena where
$(a)$ follows from the fact that $\forall \mQ\subset\mS_i$,
$R(\mQ)\ge I(X_{\mQ};Y|X_{\mS_i^c})$. The equality in $(a)$ holds if
$\mK=\mS_i\setminus\mS_{i+1}$. This proves the direct part.

To prove the converse, let $R_{\mS_1^c}\in\mR_{Y|X_{\mS_1^c}}$, and
$R_{\mS_i\setminus\mS_{i+1}}\in\mD_{YX_{\mS_i^c}|X_{\mS_i\setminus\mS_{i+1}}}$,
for $i=1,\dots,m$. We have to prove that $R(\mL)\le
I(X_{\mL};Y|X_{\mL^c})$ holds for all $\mL\subseteq [M]$ with
equality if $\mL=\mS_i$, $i=1,\dots,m$.
$R(\mS_i)=I(X_{\mS_i};Y|X_{\mS_i^c})$ is true since
$R_{\mS_i\setminus\mS_{i+1}}\in\mD_{YX_{\mS_i^c}|X_{\mS_i\setminus\mS_{i+1}}}$
and \eqa R(\mS_i)=\sum_{j=i}^m R(\mS_j\setminus\mS_{j+1})=\sum_{j=1}^m
I(X_{\mS_j\setminus\mS_{j+1}};Y|X_{\mS_j^c})=I(X_{\mS_i};Y|X_{\mS_i^c}),
\quad i=1,\dots,m.\n\ena
Now let $\mL\subseteq [M]$ and define
$\mL_i=\mL\cap\mS_i\setminus\mS_{i+1}$, $i=0,1,\dots,m$ with
$\mS_0=[M]$ by convention. Then $\mL=\bigcup_{i=0}^m \mL_i$ is a
disjoint partition. From $R_{\mS_1^c}\in\mR_{Y|X_{\mS_1^c}}$ and $\mL_0 \subseteq \mS_1^c$ it follows
$R(\mL_0)\le I(X_{\mL_0};Y|X_{\mS_1^c\setminus\mL_0})$. Furthermore, since
$R_{\mS_i\setminus\mS_{i+1}}\in\mD_{YX_{\mS_i^c}|X_{\mS_i\setminus\mS_{i+1}}}$,
$ R(\mL_i)\le
I(X_{\mL_i};Y|X_{\mS_i^c\cup(\mS_i\setminus\mS_{i+1}\setminus\mL_i)})=I(X_{\mL_i};Y|X_{\mS_{i+1}^c\setminus\mL_i})$
for all $\mL_i\subseteq\mS_i\setminus\mS_{i+1}$.
Therefore, \eqa I(X_{\mL};Y|X_{\mL^c})&=&\sum_{i=0}^m
I(X_{\mL_i};Y|X_{\bigcup_{j=0}^{i-1}\mL_j\cup\mL^c})\n\\
&\stackrel{(a)}\ge&\sum_{i=0}^m
I(X_{\mL_i};Y|X_{\mS_{i+1}^c\setminus\mL_i})\n\\
&\ge&\sum_{i=0}^mR(\mL_i)\n\\
&=&R(\mL)\n, \ena where $(a)$ holds since \eqa
\bigcup_{j=0}^{i-1}\mL_j\cup\mL^c \quad = \quad
[M]\setminus\bigcup_{j=i}^m\mL_j \quad\supseteq\quad
\mS_{i+1}^c\setminus\bigcup_{j=i}^m\mL_j \quad=\quad
\mS_{i+1}^c\setminus\mL_i,\n\ena and (b) holds since
for $j\ge i+1$, $\mL_j\subseteq\mS_{i+1}$ implies that   $\mL_j$ does not intersect with
$\mS_{i+1}^c$. This completes the proof. 

\QED

From the previous theorem, the dimension of $\mF_{\mS_1,\mS_2,...,\mS_m}$ is
\eqa \dim(\mF_{\mS_1,\mS_2,...,\mS_m}) &=& \dim(\mR_{Y|X_{\mS_1^c}})+\sum_{i=1}^{m}\dim(\mD_{YX_{\mS_i^c}|X_{\mS_i\setminus\mS_{i+1}}})\n\\
&=& M-|\mS_1|+\sum_{i=1}^{m}(|\mS_i|-|\mS_{i+1}|-1)\n\\
&=& M-m.\n\ena

The theorem also implies that all points in $\mF_{\mS_1,\mS_2,...,\mS_m}$ may be approached via group successive decoding with groups of users decoded according to the following order:
$(\mS_1^c,\mS_1\setminus\mS_2,\mS_2\setminus\mS_3,...,\mS_{m-1}\setminus\mS_m,\mS_m)$.

\bcorollary Let $\mS_1 \supset\mS_2\supset\ldots  \supset \mS_m$ be a telescopic sequence.  $R\in \mF_{\mS_1,\dots,\mS_m|\mA}$ iff $R_{\mS_1^c}\in\mR_{Y|X_{\mS_1^c}}$, $R_{\mS_i\setminus\mS_{i+1}}\in
\mD_{YX_{\mS_i^c}|X_{\mS_i\setminus\mS_{i+1}}}$ for $i=1,\dots,m$, and $R_{\mA}={\underline{0}}$. \ecorollary

$Proof:$ Recall that
$\mF_{\mS_1,\dots,\mS_m|\mA}=\mF_{\mS_1,\dots,\mS_m}\cap\mB_{\mA}$. Hence
$R\in\mF_{\mS_1,\dots,\mS_m|\mA}$
iff $R\in \mF_{\mS_1,\dots\mS_m}$ and $R_{\mA}={\underline{0}}$. The rest
follows from Theorem \ref{theo6}.

\QED

From the above Corollary we conclude that \eqa
\dim(\mF_{\mS_1,\mS_2,...,\mS_m|\mA}) =
\dim(\mF_{\mS_1,\mS_2,...,\mS_m})-|\mA| = M-|\mA|-m.\n\ena
Furthermore,  $R\in \mF_{\mS_1,\dots,\mS_m|\mA}$  may be approached by decoding groups of users in the order $([M]\setminus\mA\setminus\mS_1,\mS_1\setminus\mS_2,\mS_2\setminus\mS_3,...,\mS_{m-1}\setminus\mS_m,\mS_m)$.

\section{Number of faces of dimension $D$}
\label{numdfac}

Now we are ready to derive the number of $D$-dimensional faces in
$\mR$ for any $D=0,1,\dots,M$. We start by describing the number of
$D$-dimensional faces of the dominant facet.

\bproposition The number of $D$-dimensional faces in the dominant facet of $\mR$ is
\eqa
N_d(M,D)&=&\sum_{j=1}^{M-D}\binom{M-D}{j}(-1)^{M-D-j}j^M. \label{2red}\ena
\label{teoD}\eproposition
{\it Proof:} Any $D$-dimensional face on
the dominant facet is labeled by $\mF_{[M],\mS_2,\dots,\mS_{M-D}}$.
The difference sets
$[M]\setminus\mS_{2},\mS_2\setminus\mS_{3},\dots,\mS_i\setminus\mS_{i+1},\dots,
\mS_{M-D}\setminus\emptyset$ form an $(M-D)$ partition of $[M]$.
There is a one-to-one correspondence between a $D$-dimensional face
and such a partition. The number of such ordered partitions is
\begin{align}
N_d(M,D)=\sum_{\substack{m_1,m_2,\dots,m_{M-D}\\m_i\ge
1,\forall i\\
\sum_im_i=M}}\binom{M}{m_1,m_2,\dots ,m_{M-D}} = \sum_{\substack{m_1,m_2,\dots,m_{M-D}\\m_i\ge 1,\forall i\\
\sum_im_i=M}}\frac{M!}{\prod_{i}^{}m_i!}.\label{ddq}
\end{align}
To go further,  we expand the following polynomial 
\eqa
\left(\frac{x}{1!}+\frac{x^2}{2!}+\dots+\frac{x^M}{M!}\right)^{M-D}
=\sum_{k=M-D}^{M(M-D)}x^k\sum_{\substack{m_1,\dots,m_{M-D}\\m_i\ge
1,\forall i\\ \sum_im_i=k}}\frac{1}{m_1!m_2!\dots m_{M-D}!},\n 
\ena 
and note that the coefficient in front
of $x^M$ multiplied by $M!$ gives (\ref{ddq}). Therefore, \eqa
N_d(M,D)&=&M!\coeff\left(\left(\sum_{i=1}^{M}\frac{x^i}{i!}\right)^{M-D},x^M\right)\n\\
&\stackrel{(a)}=&M!\coeff\left(\left(\sum_{i=1}^{\infty}\frac{x^i}{i!}\right)^{M-D},x^M\right)\n\\
&\stackrel{(b)}=&M!\coeff\left((e^x-1)^{M-D},x^M\right)\n\\ 
&\stackrel{(c)}=&\left.\frac{d^{M}}{dx^M}(e^x-1)^{M-D}\right|_{x=0},\n\ena
where coeff$(f(x),x^i)$ is the coefficient of $x^i$ in the Taylor
series expansion around zero of the function $f(x)$, $(a)$ is true
since taking all the terms up to $M$ or
up to infinity will not change the coefficient in front of $x^M$, $(b)$ follows from the Taylor expansion of $e^x$,  and $(c)$ follows
from the definition of the Taylor expansion.

To prove (\ref{2red}), we use the Binomial formula to expand $(e^x-1)^{M-D}$, namely 
\eqa (e^x-1)^{M-D}=\sum_{j=0}^{M-D}\binom{M-D}{j}
e^{jx}(-1)^{M-D-j}.\n\ena Taking the $M$-th derivative,
\eqa\frac{d^M
}{dx^M}(e^x-1)^{M-D}=\sum_{j=1}^{M-D}\binom{M-D}{j}
e^{jx}(-1)^{M-D-j}j^M,\n\ena and setting $x=0$ we obtain (\ref{2red}).

\QED

Observe that by letting $D=0$, using the fact that there are $M!$
vertices in the dominant facet, from (\ref{2red}) we obtain an
alternative expression for $M!$ that is \eqa
M!=\sum_{j=1}^{M}\binom{M}{j}(-1)^{M-j}j^M.\n\ena

The number of faces in the dominant facet is directly connected to the {\it Stirling number of the second kind} \cite{stir,wilf}, denoted by $\stir{M}{n}$. This is known as {\it Karamata notation} \cite{karamata}. The Stirling number of the second kind is the number of ways we can  partition a set of $M$ elements into $n$ nonempty subsets. In  calculating the number of $D$-dimensional faces, all permutations of such partitions have to be counted. That is,
\eqa N_d(M,D)=(M-D)!\stir{M}{M-D}.\n\ena

Like for facets, it is useful to distinguish between front and back faces. Hence we say that a face $\mF_{\mS_1,\mS_2,\dots,\mS_{m}|\mA}$, is called a {\it
front face} if $\mA=\emptyset$ and  a {\it back face} if
$\mA\ne\emptyset$.

\bproposition
 The total number of front faces of dimension
 $D$, denoted by $N_f(M,D)$, equals
\eqa N_f(M,D)=N_d(M,D)+N_d(M,D-1).\label{wqw}\ena
\label{prop1}\eproposition 
{\it Proof:} Any $D$-dimensional front
face has a label $\mF_{\mS_1,\mS_2,\dots,\mS_{M-D}|\emptyset}$ for
some $\mS_1\subseteq [M]$. If $\mS_1=[M]$ , the front face
is in the dominant facet and there are $N_d(M,D)$ such faces. If
$\mS_1\subset [M]$ the front face is not in the dominant
facet. Since there is a one-to-one relationship between the
subscripts of $\mF_{\mS_1,\mS_2,\dots,\mS_{M-D}|\emptyset}$ and those of
$\mF_{[M],\mS_1,\mS_2,\dots,\mS_{M-D}|\emptyset}$ when $\mS_1\subset
[M]$, if follows that the total number of front faces not in the
dominant facet is exactly $N_d(M,D-1)$. To obtain the total
number of front $D$-faces we have to add this number and the number $N_d(M,D)$
of $D$-faces in the dominant facet.  

\QED

We now have an expression for $N_d(M,D)$ (Proposition \ref{teoD})
and an expression for $N_f(M,D)$ (Proposition \ref{prop1}). Next we
derive an expression for the number of back faces $N_b(M,D)$.

\bproposition The total number of $D$-dimensional back faces in $\mR$ is given by \eqa
N_b(M,D)=\sum_{i=D}^{M-1}\binom{M}{i}N_f(i,D). \label{bafac} \ena
\eproposition

 {\it Proof:} To derive (\ref{bafac}), we observe that
all back faces are front faces for some  other channel  with fewer users. This can
be  seen from the label ${\cal
F}_{\mS_1,\mS_2,\dots,\mS_{m}|\mA}$ of a  back face, where $\mA\ne\emptyset$. The dimension
of this face is  $M-m-|\mA|$. Recall that $\mA\cap\mS_1=\emptyset$. If
we remove all users with index in $\mA$, we obtain the front face
${\cal F}_{\mS_1,\mS_2,\dots,\mS_{m}|\emptyset}$ of an  $(M-|\mA|)$-user MAC.
The dimensionality of this face is also
$M-|\mA|-m$. Running over all pertinent subsets  $\mA\subset [M]$ yields
\eqa N_b(M,D)=\sum_{\substack{\mA\subset [M]\\ 0<|\mA|\le
M-D}}N_f(M-|\mA|,D).\n\ena
Since there are $\binom{M}{|\mA|}$ subsets of cardinality $|\mA|$,
\eqa N_b(M,D)&=&\sum_{|\mA|=1}^{M-D} \binom{M}{|\mA|}N_f(M-|\mA|,D)=\sum_{i=1}^{M-D} \binom{M}{i}N_f(M-i,D)\n\\
&=&\sum_{i=1}^{M-D} \binom{M}{M-i}N_f(M-i,D)=\sum_{i=D}^{M-1} \binom{M}{i}N_f(i,D).\n\ena
 
 \QED

Now we are ready to derive an expression for the total number of
$D$-dimensional faces in $\mR$.

\btheorem The total number of $D$-dimensional faces in $\mR$, $0\le
D\le M$, is
\eqa
N(M,D)=\sum_{i=D}^{M}\binom{M}{i}\left[(i+1-D)^i-\sum_{j=1}^{i-D}\binom{i-D}{j-1}(-1)^{i-D-j}j^i\right].\label{finalresult} \ena
 \etheorem

{\it Proof:} First we observe that \eqa
N(M,D)=N_f(M,D)+N_b(M,D)=\sum_{i=D}^{M}\binom{M}{i}N_f(i,D).\n\ena
Using (\ref{wqw}) we obtain \eqa
N(M,D)=\sum_{i=D}^M \binom{M}{i}
[N_d(i,D)+N_d(i,D-1)],\label{sumdf}\ena where $N_d(D,D)=0$,
$N_d(D,D-1)=1$ and, by convention, $N_d(i,-1)=0$ (the latter is needed for the case $D=0$).
Furthermore, from (\ref{2red})
we obtain \eqa
&\quad&N_d(i,D)+N_d(i,D-1)\n\\
&=&(i-D+1)^i+\sum_{j=0}^{i-D}j^i (-1)^{i-D-j+1} \left[\binom{i-D+1}{j}-\binom{i-D}{j}\right]\n\\
&=&(i-D+1)^i-\sum_{j=0}^{i-D}\binom{i-D}{j}\frac{j^{i+1}
(-1)^{i-D-j}}{i-D+1-j}\n\\
&=&(i-D+1)^i-\sum_{j=1}^{i-D}\binom{i-D}{j-1}(-1)^{i-D-j}j^i.
\label{last}\ena
Inserting (\ref{last}) into (\ref{sumdf}) yields (\ref{finalresult}) and completes the proof.  

\QED

Next we determine a closed form expression for the total number $N(M,0)$ of
vertices and the total number $N(M,1)$ of edges.

\blemma \label{lemver} The total number of vertices in $\mR$ is $\lfloor
eM!\rfloor$. \elemma

{\it Proof:} From (\ref{sumdf}), \eqa
N(M,0)&=&\sum_{i=0}^M\binom{M}{i}N_d(i,0)\n\\
&\stackrel{(a)}{=}&\sum_{i=0}^M\binom{M}{i}i!=\sum_{i=0}^M\frac{M!}{(M-i)!}=\sum_{i=0}^M\frac{M!}{i!}\n\\
&=&M!\sum_{i=0}^{\infty}\frac{1}{i!}-M!\sum_{i=M+1}^{\infty}\frac{1}{i!}\n\\
&\stackrel{(b)}{=}&eM!-M!\sum_{i=M+1}^{\infty}\frac{1}{i!},\n\ena
where in $(a)$ we have used the well known fact that the number of
vertices $N_d(i,0)$ of the dominant facet of an $i$-user
region is $i!$ and $(b)$ follows from the Taylor series expansion of
$e$. Since $eM!-M!\sum_{i=M+1}^{\infty}\frac{1}{i!}$ is an integer,
and \eqa
\sum_{i=M+1}^{\infty}\frac{M!}{i!}&=&\sum_{i=1}^{\infty}\frac{M!}{(M+i)!}
=\sum_{i=1}^{\infty}\frac{1}{\prod_{j=1}^i(M+j)}\n\\
&<&\sum_{i=1}^{\infty}\prod_{j=1}^i\frac{1}{M+1}=\sum_{i=1}^{\infty}\left(\frac{1}{M+1}\right)^i\n\\
&=&\frac{1/(M+1)}{1-1/(M+1)}=\frac{1}{M}\le 1, \n\ena it follows that
\eqa N(M,0)=\sum_{i=0}^M\frac{M!}{i!}=\left\lfloor
N(M,0)+\sum_{i=M+1}^{\infty}\frac{M!}{i!}\right\rfloor=\lfloor eM!
\rfloor.\label{kkn}\ena 

\QED

\blemma \label{Ned_lema} The total number of edges in $\mR$  is $\dfrac{M}{2}\lfloor eM! \rfloor$.
\elemma

{\it Proof:} From (\ref{sumdf}) we have \eqa N(M,1)=\sum_{i=1}^M\binom{M}{i}(N_d(i,1)+N_d(i,0)).\n\ena Furthermore, since $N_d(i,0)=i!$, from (\ref{ddq}),
 \eqa N_{d}(i,1)=\sum_{\substack{m_1,m_2,\dots,m_{i-1}\\m_j\ge
1,\forall j\\ \sum_jm_j=i}}\binom{i}{m_1,\dots
,m_{i-1}}=(i-1)\binom{i}{2,1,\dots,1}=\frac{i!(i-1)}{2}.\n\ena
 Therefore,
\eqa
N(M,1)&=&\sum_{i=1}^{M} \binom{M}{i}\left(i!+\frac{i-1}{2}i!\right)
=\frac{1}{2}\sum_{i=1}^{M} \binom{M}{i}i!\left(i+1 \right)\n\\
&=&\frac{1}{2}\sum_{i=1}^{M} \frac{M!}{(M-i)!}\left(i+1
\right)=\frac{1}{2}\sum_{j=0}^{M-1} \frac{M!}{j!}\left(M-j+1
\right)\n\\
&=&\frac{M+1}{2}\sum_{j=0}^{M-1}
\frac{M!}{j!}-\frac{1}{2}\sum_{k=0}^{M-2}
\frac{M!}{k!}\n\\
&\stackrel{(a)}{=}&\frac{1}{2}[(M+1)(\lfloor eM! \rfloor-1)-(\lfloor eM! \rfloor-M-1)]=\frac{M}{2}\lfloor eM! \rfloor,\n\ena  where in $(a)$ we use
(\ref{kkn}) to obtain $\sum_{j=0}^{M-1} M!/j!=\lfloor eM!
\rfloor-1$ and $\sum_{j=0}^{M-2} M!/j!=\lfloor eM! \rfloor-M-1$.

\QED


\section{Summary}
\label{con}

The capacity region of an asynchronous memoryless multiple-access channel is the union of certain polytopes. The points in those polytopes are exactly the rate tuples that can be approached at an arbitrarily small error probability.
In this paper we have developed operational and  structural properties that apply to  those polytopes. The centerpiece of our developments are the labels that we use to tag their faces. For non-degenerated cases (the only kind considered in this paper), the set of labels is the set of expressions of the form $(\mS_1,\mS_2,\ldots,\mS_m|\mA)$, where $\mA\subseteq[M]$ and  $[M]\setminus\mA\coc \mS_1 \coc \mS_2,\ldots,\coc\mS_m$. This extends the labeling introduced in \cite{ru99}.
Each label of the above form tags one face and each face has a unique such tag.
We have shown that the  label  $\mS_1,\mS_2,\ldots,\mS_m|\mA$ tags a face of dimension $M-m-|\mA|$. By counting the number of such expressions for a fixed $k$, we find the number of faces of a given dimension.

We have also shown that a rate tuple on the face with  label $\mS_1,\mS_2,\ldots,\mS_m|\mA$ may be approached via successive decoding, as follows: the users with index in
$([M]\setminus\mA\setminus\mS_1$ are decoded first, followed by the users with index in $\mS_1\setminus\mS_2$, followed by those with index in $\mS_2\setminus\mS_3$ etc. The users with index in $\mS_m$ are decoded last. The users with index in $\mA$ do not need to be decoded since they have vanishing rate. The decoding order $([M]\setminus\mA\setminus\mS_1,\mS_1\setminus\mS_2,\mS_2\setminus\mS_3,...,\mS_{m-1}\setminus\mS_m,\mS_m)$ is an equivalent alternative way to label faces.

\begin{figure}[ht]
\begin{center}
\leavevmode
\includegraphics[width=12cm]{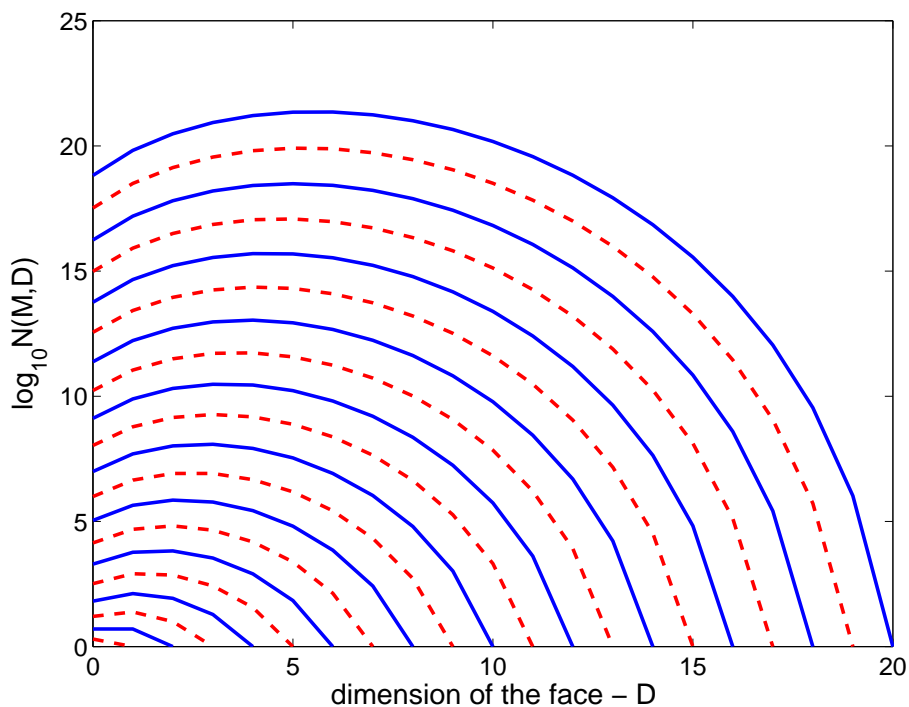}
\end{center}
\caption{Total number of $D$-dimensional
faces (expressed in logarithmic form) as a function of $D$. Each curve corresponds to a value of $M$.   The curve that corresponds to
$M=m$, $m=1,2,\dots,20$, is the one that hits the abscissa at $D=m$.} \label{totDofM}.
\end{figure}

Table \ref{tab61} summarizes the expressions for the number of faces of a given dimension, where $f_n^{(i)}(0)=\left.\frac{d}{dx}(e^x-1)^n\right|_{x=0}$. The logarithm of the total number of $D$-dimensional faces as a
function of $D$, for $M=1,2,\dots,20$ is shown in
Fig.~\ref{totDofM}.

\begin{table}[ht]
\caption{Number of vertices, edges, facets and $D$-dimensional
faces for an $M$-user MAC.}
\begin{center}
\begin{tabular}{{p{2.5cm}  p{7cm}  p{4.0cm}}}
\hline\hline
\quad & & \\
{Objects} & In $\mR$  & In the dominant facet\\[2.0ex]
\hline 
\quad & & \\
Vertices& $\lfloor eM!\rfloor$ & $M!$\\[1.5ex]
Edges& $\frac{M}{2}\lfloor eM!\rfloor$ & $M!(M-1)/2$\\[1.5ex]
Facets& $M+2^M-1$ & $2^M-2$\\[1.5ex]
$D$-faces& $\sum_{i=D}^M\binom{M}{i}\left(f_{i-D}^{(i)}(0)+f_{i-D+1}^{(i)}(0)\right)$ & $f_{M-D}^{(M)}(0)$\\[1.5ex]
\hline
\end{tabular}
\end{center}
\label{tab61}
\end{table}

 \end{document}

%% file: degenerated2.pstex_t
\begin{picture}(0,0)%
\includegraphics{degenerated2.pstex}%
\end{picture}%
\setlength{\unitlength}{3552sp}%
\begingroup\makeatletter\ifx\SetFigFont\undefined%
\gdef\SetFigFont#1#2#3#4#5{%
  \reset@font\fontsize{#1}{#2pt}%
  \fontfamily{#3}\fontseries{#4}\fontshape{#5}%
  \selectfont}%
\fi\endgroup%
\begin{picture}(6184,2734)(879,-2633)
\put(1951,-1036){\makebox(0,0)[lb]{\smash{{\SetFigFont{9}{10.8}{\rmdefault}{\mddefault}{\updefault}{\color[rgb]{0,0,0}$R_1$}%
}}}}
\put(976,-61){\makebox(0,0)[lb]{\smash{{\SetFigFont{9}{10.8}{\rmdefault}{\mddefault}{\updefault}{\color[rgb]{0,0,0}$R_2$}%
}}}}
\end{picture}%

%% file: cr3Ddegenerated2.pstex_t
\begin{picture}(0,0)%
\includegraphics{cr3Ddegenerated2.pstex}%
\end{picture}%
\setlength{\unitlength}{3552sp}%
\begingroup\makeatletter\ifx\SetFigFont\undefined%
\gdef\SetFigFont#1#2#3#4#5{%
  \reset@font\fontsize{#1}{#2pt}%
  \fontfamily{#3}\fontseries{#4}\fontshape{#5}%
  \selectfont}%
\fi\endgroup%
\begin{picture}(7033,5250)(35,-5179)
\put(451,-1486){\makebox(0,0)[rb]{\smash{{\SetFigFont{9}{10.8}{\rmdefault}{\mddefault}{\updefault}{\color[rgb]{0,0,0}$R_1$}%
}}}}
\put(1951,-886){\makebox(0,0)[lb]{\smash{{\SetFigFont{9}{10.8}{\rmdefault}{\mddefault}{\updefault}{\color[rgb]{0,0,0}$R_2$}%
}}}}
\put(1126,-61){\makebox(0,0)[lb]{\smash{{\SetFigFont{9}{10.8}{\rmdefault}{\mddefault}{\updefault}{\color[rgb]{0,0,0}$R_3$}%
}}}}
\end{picture}%

%% file: cr3Dbest.pstex_t
\begin{picture}(0,0)%
\includegraphics{cr3Dbest.pstex}%
\end{picture}%
\setlength{\unitlength}{2486sp}%
\begingroup\makeatletter\ifx\SetFigFont\undefined%
\gdef\SetFigFont#1#2#3#4#5{%
  \reset@font\fontsize{#1}{#2pt}%
  \fontfamily{#3}\fontseries{#4}\fontshape{#5}%
  \selectfont}%
\fi\endgroup%
\begin{picture}(6656,6181)(3072,-6602)
\put(6001,-589){\makebox(0,0)[lb]{\smash{{\SetFigFont{10}{12.0}{\rmdefault}{\mddefault}{\updefault}{\color[rgb]{0,0,0}$R_3$}%
}}}}
\put(3144,-6304){\makebox(0,0)[lb]{\smash{{\SetFigFont{10}{12.0}{\rmdefault}{\mddefault}{\updefault}{\color[rgb]{0,0,0}$R_1$}%
}}}}
\put(4930,-3518){\makebox(0,0)[b]{\smash{{\SetFigFont{10}{12.0}{\rmdefault}{\mddefault}{\updefault}{\color[rgb]{0,0,0}$\mF_{\{1,3\}}$}%
}}}}
\put(6073,-2161){\makebox(0,0)[b]{\smash{{\SetFigFont{10}{12.0}{\rmdefault}{\mddefault}{\updefault}{\color[rgb]{0,0,0}$\mF_{\{3\}}$}%
}}}}
\put(5073,-5233){\makebox(0,0)[b]{\smash{{\SetFigFont{10}{12.0}{\rmdefault}{\mddefault}{\updefault}{\color[rgb]{0,0,0}$\mF_{\{1\}}$}%
}}}}
\put(6787,-5304){\makebox(0,0)[b]{\smash{{\SetFigFont{10}{12.0}{\rmdefault}{\mddefault}{\updefault}{\color[rgb]{0,0,0}$\mF_{\{1,2\}}$}%
}}}}
\put(8144,-4161){\makebox(0,0)[b]{\smash{{\SetFigFont{10}{12.0}{\rmdefault}{\mddefault}{\updefault}{\color[rgb]{0,0,0}$\mF_{\{2\}}$}%
}}}}
\put(7716,-2732){\makebox(0,0)[b]{\smash{{\SetFigFont{10}{12.0}{\rmdefault}{\mddefault}{\updefault}{\color[rgb]{0,0,0}$\mF_{\{2,3\}}$}%
}}}}
\put(6858,-1232){\makebox(0,0)[b]{\smash{{\SetFigFont{10}{12.0}{\rmdefault}{\mddefault}{\updefault}{\color[rgb]{0,0,0}$\mF_{\{1,2,3\},\{2,3\},\{3\}}$}%
}}}}
\put(9716,-4090){\makebox(0,0)[rb]{\smash{{\SetFigFont{10}{12.0}{\rmdefault}{\mddefault}{\updefault}{\color[rgb]{0,0,0}$R_2$}%
}}}}
\put(8573,-2161){\makebox(0,0)[b]{\smash{{\SetFigFont{10}{12.0}{\rmdefault}{\mddefault}{\updefault}{\color[rgb]{0,0,0}$\mF_{\{2,3\},\{2\}|\{1\}}$}%
}}}}
\put(3429,-3875){\makebox(0,0)[rb]{\smash{{\SetFigFont{10}{12.0}{\rmdefault}{\mddefault}{\updefault}{\color[rgb]{0,0,0}$\mF_{\{1,2,3\},\{1,3\}}$}%
}}}}
\put(3072,-5161){\makebox(0,0)[rb]{\smash{{\SetFigFont{10}{12.0}{\rmdefault}{\mddefault}{\updefault}{\color[rgb]{0,0,0}$\mF_{\{1\}|\{2\}}$}%
}}}}
\put(6716,-3590){\makebox(0,0)[b]{\smash{{\SetFigFont{10}{12.0}{\rmdefault}{\mddefault}{\updefault}{\color[rgb]{0,0,0}\textsc{Dominant Facet}}%
}}}}
\put(6716,-3875){\makebox(0,0)[b]{\smash{{\SetFigFont{10}{12.0}{\rmdefault}{\mddefault}{\updefault}{\color[rgb]{0,0,0}$\mF_{\{1,2,3\}}$}%
}}}}
\end{picture}%